\documentclass[10pt,twocolumn,twoside]{IEEEtran}
\newcommand{\comm}[1]{}
\usepackage{amssymb}
\usepackage{amsthm}
\usepackage{etoolbox}
  \usepackage{epsfig}
  \usepackage{color}

\usepackage[numbers]{natbib}
\def\citet{\cite}

   \renewcommand{\theequation}{\arabic{equation}}

   \renewcommand{\theequation}{\thesection.\arabic{equation}}
   \csname @addtoreset\endcsname{equation}{section}
\newcounter{thanksnum}
\def\thanksnumber#1
{\setcounter{thanksnum}{\value{footnote}}\setcounter{footnote}{#1}%
                     \addtocounter{footnote}{-1}\footnotemark
                     \setcounter{footnote}{\value{thanksnum}}}

\newtheorem{theorem}{Theorem}
\newtheorem{lemma}{Lemma}
\newtheorem{proposition}{Proposition}
\newtheorem{corollary}{Corollary}
\newtheorem{remark}{Remark}

\def\defi{\stackrel{{\scriptscriptstyle \Delta}}{=}}

\def\o{\omega}
\def\O{\Omega}

\def\Y{{\cal Y}}

\def\w{\widehat}
\def\Ind{{\,\rm Ind\,}}
\def\Ind{{\mathbb{I}}}

\def\R{{\bf R}}
\def\E{{\bf E}}

\def\Z{{\cal Z}}

\def\H{{\cal H}}

\def\C{{\bf C}}

\def\ww{\widetilde}
\def\X{{\cal X}}
\def\t{\theta}
\def\oo{\bar}

\newcommand{\be}{\begin{equation}}
\newcommand{\ee}{\end{equation}}
\newcommand{\bd}{\begin{displaymath}}
\newcommand{\ed}{\end{displaymath}}
\newcommand{\ba}{\begin{array}{ll}}
\newcommand{\ea}{\end{array}}
\newcommand{\baa}{\begin{eqnarray}}
\newcommand{\eaa}{\end{eqnarray}}
\newcommand{\baaa}{\begin{eqnarray*}}
\newcommand{\eaaa}{\end{eqnarray*}}
\font\sm=cmr10

\def\BS{{\scriptscriptstyle BS}}
\def\oo{\bar}

\def\BL{\b_{\scriptscriptstyle min}}
\def\BL{{\scriptscriptstyle BL}}
\def\NS{{\scriptscriptstyle N}}
\def\None{{\scriptscriptstyle N_1}}
\def\Ntwo{{\scriptscriptstyle N_2}}

\def\sinc{{\rm sinc\,}}
\def\ew{\left(e^{i\o}\right)}
\def\T{{\mathbb{T}}}
\def\ZZ{{\mathbb{Z}}}

\def\TT{{\cal \ZZ^-}}
\def\XN{\ell_2^\BL}
\def\XNL{\ell_2^\BL(-\infty,0)}
\def\BN{{L_2^\BL(\T)}}
\def\BNN{{L_2^\BL(\T)}}
\def\nnn{\left\|(I-A_\rho)^{-1}\right\| }
\def\nnn{(1+\rho^{-1})}
\def\EE{\mathbb{E}}
\date{Submitted April 9, 2015. Revised: September 24, 2017}
\title{A closed equation in time domain for band-limited extensions of one-sided sequences}
\author{
Nikolai Dokuchaev }
\begin{document}
\def\break{}%
\def\brea{}
\def\breakk{}
\def\brea{\nonumber\\ }\def\breakk{\nonumber\\&&} 
\maketitle
\let\thefootnote\relax\footnote{Accepted to {\em IEEE Transactions on Signal Processing}.}
\let\thefootnote\relax\footnote{This work was supported by ARC grant of Australia
DP120100928 to the author. \par The author is with the
 Department of Mathematics \& Statistics, Curtin
University, GPO Box U1987, Perth, 6845 Western Australia,  and also with National Research University ITMO, 197101 Russia  (email N.Dokuchaev@curtin.edu.au)}
\begin{abstract}  The paper suggests  a  method of optimal  extension
of one-sided semi-infinite sequences of a general type by traces of band-limited
sequences in deterministic setting, i.e.  without probabilistic assumptions. The method requires to solve
a closed linear equation in the time domain connecting
the past observations of the underlying process with the future values of the band-limited process.
 Robustness of the solution with respect to the input errors and data truncation
is established  in the framework of Tikhonov regularization.
\end{abstract}
\section{Introduction}  We study extrapolation
of one-sided semi-infinite sequences in pathwise deterministic setting.
 Extrapolation of sequences can be used for forecasting and was
studied intensively, for example,  in the framework of system identification methods; see e.g. \cite{SFS}.
In signal processing, there is a different approach oriented on the
frequency analysis and exploring special features of the band-limited processes such as a  uniqueness of extrapolation.
The present paper extends this approach
on  processes that are not necessarily band-limited; we consider
 extrapolations of the optimal  band-limited approximations of the
observed parts  of underlying processes.
The motivation for that approach is based on the premise that a band-limited approximation of a process can be interpreted as its regular part purified
from a noise represented by the high-frequency component.
This leads to a problem of causal band-limited approximations for non-bandlimited
underlying processes.
In theory, a process can be converted into a band-limited process with a low-pass filter, and the resulting process will
be an optimal band-limited approximation.    However,
 a ideal low-pass filter is non-causal; therefore, it cannot be applied for a process that is observable
 dynamically such that its future values are unavailable which is crucial for predicting and extrapolation problems.
 It is known that the distance of an ideal low-pass
filter from the set of all causal filters is positive \cite{rema}.
Respectively,  causal smoothing cannot convert a process into a band-limited one.
 There are many works devoted to causal smoothing and sampling, oriented on estimation and minimization of errors in $L_2$-norms or similar norms, especially in stochastic setting; see e.g.
\citet{Alem,CTao,CJR,D16,D17,PFG,jerri,K,W,Zhao, Zhao2}.
  \index{\citet{jerri}, \citet{PFG}, \citet{PFG1}, \citet{AU}.} 

The present paper considers the problem of causal  band-limited extrapolation for one-sided semi-infinite sequences that
are not are not necessarily traces of band-limited processes. We  consider purely discrete time processes rather than samples of continuous time processes.
This setting imposes certain restrictions. In particular, it does not allow to consider continuously
variable locations of the sampling points, as is common in sampling analysis of continuous time processes;
see e.g. \citet{BS,PFG,FKR,LeeF}. In our setting,
the values between fixed discrete times are  not included into consideration. For continuous
time processes, the predicting horizon can be selected to be arbitrarily  small, such as in the model considered  in \citet{BS};
this possibility is absent for discrete time processes considered below.

Further, we consider the extrapolation problem  in the pathwise deterministic setting,  without probabilistic assumptions.
This means that the method has to rely on the intrinsic properties of a sole underlying sequence without appealing
to statistical properties of an ensemble  of sequences. In particular, we use a
pathwise optimality criterion rather than criterions calculated via the expectation on a probability space such as
mean variance criterions.

In addition, we consider an approximation that does not target the match of the values at any set of selected points; the error is not expected to be small.
This is different from  a more common setting where the goal is to match
an approximating curve with the underlying process
at certain sampling points; see e.g. \citet{CJR,FKR,jerri,LeeF,S78,Ka}.
Our setting is closer to the setting from \citet{PFG,PFG1,TH,Zhao,Zhao2}.
In \citet{PFG,PFG1}, the point-wise matching error
was estimated for a sampling series and for a band-limited process representing
smoothed
underlying  continuous time process; the  estimate featured a
given vanishing error. In \citet{TH}, the problem  of minimization of the total energy  of the approximating bandlimited process was considered; this causal approximation was constructed
 within  a given distance from  the original process smoothed by an ideal low-pass filter.
 Another related result was obtained in \citet{F94}, where an interpolation problem for absent sampling points  was considered in a setting with vanishing error, for a finite number of sampling points.
 In \cite{S78,Zhao,Zhao2,Ka}, extrapolation of a trace of a band-limited process
 was investigated using some special Slepian's type basis \citet{SP,S78}  in the frequency domain.
  In \cite{S78}, the idea of this extrapolation was suggested as an example of applications of this basis.
  In \citet{Zhao},  extrapolation of a trace of a band-limited process from a finite number of points was considered in a frequency
 setting for a general linear transform and some special Slepian's type basis \citet{SP,S78}  in the frequency domain. In \citet{Zhao2}, a setting  similar to \citet{Zhao}
 was considered for extrapolation of a trace of continuous time process from a finite interval using a special basis from eigenfunctions  in the frequency domain. In \cite{Ka},  extrapolation of a trace of a band-limited process
 was considered as an example of applications for a numerically efficient version of the Slepian basis.
 Our setting is different:
we consider extrapolation in time domain. The paper offers a new method of calculating the future values of the optimal band-limited approximation, i.e. the extrapolation of the
approximating trace of an optimal band-limited process on the future times.
 The underlying process does not have to be a trace of a band-limited process;  therefore,  there is a non-vanishing approximation error being  minimized.
The problem is reduced to solution of a convenient closed linear equation  connecting directly
the set of past observations of the underlying process with the set of future values of the band-limited process
(equation (\ref{yAa}) in Theorem \ref{ThM} and equation (\ref{yAaR}) in Theorem \ref{ThMR} below).
This allows to bypass analysis in the frequency domain and skip calculation
of the past values for the  approximating band-limited process; respectively, a non-trivial procedure of
 extrapolation of a band-limited process from its part is also bypassed. This  streamlines
the calculations. We study this equation  in the time domain, without transition to the frequency domain; therefore,
the selection of the basis in the frequency domain
is not required.
  We established  solvability and uniqueness of the solution of the suggested equation for the band-limited extension.
Furthermore,  we established  numerical stability and robustness of the method  with respect to the input errors and data
 truncation in a version of the problem where there is a penalty on the norm of the approximating band-limited process, i.e. under  Tikhonov regularization (Theorem \ref{ThMR}).
 We found that this regularization can be achieved
   with  an arbitrarily small modification of the optimization problem.

 We illustrated  the  sustainability
   of the  method with  some  numerical experiments where we compare the band-limited extrapolation with some classical spline based interpolations (Section \ref{secN}).

 \section{Some definitions and background}

Let $\ZZ$  be the set of all integers, let $\ZZ^+=\{1,2,3,...\}$, and let $\ZZ^-=\{...,-3,-2,-1,0\}$. \par

We denote by $\ell_2(\t,\tau)$ a Hilbert
space of real valued sequences $\{x(t)\}_{t=\t}^\tau$
such that
$\|x\|_{\ell_2(\t,\tau)}=\left(\sum_{t=\t}^\tau|x(t)|^2\right)^{1/2}<+\infty$.

Let $\ell_2=\ell_2(-\infty,+\infty)$,  \index{let $\ell_2^-$ be the subspace in $\ell_2$ consisting of all
$x\in\ell_2$ such that $x(t)=0$ for $t\ge 0$, and} and let $\ell_2^+$ be the subspace in $\ell_2$ consisting of all
$x\in\ell_2$ such that $x(t)=0$ for $t<0$.
\par
For  $x\in \ell_2$, we denote by $X=\Z x$ the
Z-transform  \baaa X(z)=\sum_{t=-\infty}^{\infty}x(t)z^{-t},\quad
z\in\C. \eaaa Respectively, the inverse $x=\Z^{-1}X$ of the Z-transform   is
defined as \baaa x(t)=\frac{1}{2\pi}\int_{-\pi}^\pi
X\left(e^{i\o}\right) e^{i\o t}d\o, \quad t=0,\pm 1,\pm 2,....\eaaa

We assume that we are given $\O\in(0,\pi)$.

\par
Let $\T=\{z\in\C:\ |z|=1\}$.
\par
Let $\BNN$ be the set of all mappings $X:\T\to\C$ such
that $X\ew \in L_2(-\pi,\pi)$ and $X\ew =0$ for $|\o|>\O$, $\o\in (-\pi,\pi]$. We will call  the corresponding processes $x=\Z^{-1}X$
{\em band-limited}.

Let $\XN$ be the set of all band-limited processes from $\ell_2$, and let $\XNL$ be the subset of $\ell_2(-\infty,0)$ formed by the traces
$\{\w x(t)\}_{t\le 0}$ for all sequences  $\w x\in\XN$.

We will use the notation $\sinc(x)=\sin(x)/x$, and we will  use the notation ``$\circ$''  for the convolution in $\ell_2$.

Let $H(z)$ be the transfer function for an ideal low-pass filter such that $H\ew=\Ind_{[-\O,\O]}(\o)$, where
$\Ind$ denotes the indicator function, $\o\in (-\pi,\pi]$. Let $h=\Z^{-1}H$;
it is known that  $h(t)=\O\,\sinc(\O t)/\pi$.
The  definitions imply that $h\circ x\in \XN$ for any $x\in \ell_2$.


\begin{proposition}\label{propU}  For
  any $x\in\XNL$, there exists a unique $\w x\in\XN$ such that $\w x(t)=x(t)$ for $t\le 0$.
\end{proposition}
\par
Proposition \ref{propU} implies that the future $\{\w x(t)\}_{t>0}$ of a
band-limited process
 is uniquely defined by its  past
$\{\w x(t),\ t\le 0\}$. This can be considered as  reformulation in the deterministic setting
of a sufficient condition of predictability  implied by the classical Szeg\"o-Kolmogorov Theorem  for stationary Gaussian processes
\citet{K,Sz,Sz1}; more recent review can be found in \citet{Bin,S}.

\section{The main results}
We consider below input processes $x\in\ell_2(-\infty,0)$ and their band-limited approximations and extensions.
The sequences $\{x(t)\}_{t\le 0}$ represent the historical data available at the current time $t=0$;
the future values for $t>0$ are unavailable.

\subsection{Existence and uniqueness of the band-limited extension}
Clearly, it is impossible  to apply the ideal low-pass filter  directly to the underlying processes $x\in\ell_2(-\infty,0)$  since
the convolution with $h$ requires the future values that are unavailable. We will
be using approximation described in the following lemma.
\begin{lemma}\label{lemma1} There exists a unique optimal solution  $\w x\in\XN$
of the minimization problem \baa &&\hbox{Minimize}\quad  \sum_{t\le 0 }|
x_\BL(t)-x(t)|^2  \quad\breakk\hbox{over}\quad x_\BL\in \XN .\label{min} \eaa
\end{lemma}

\par
Under the assumptions of Lemma \ref{lemma1},
there exists a unique band-limited process $\w x$ such that its trace $\w x|_{t\le 0}$
provides an optimal approximation of
the observable past path  $\{ x(t)\}_{t\le 0}$.  The corresponding  future path  $\{\w x(t)\}_{t> 0}$  can
be interpreted  as an optimal forecast of $x$ (optimal in the sense of problem (\ref{min}) given $\O$). We will suggest below a method of calculation of
this future path $\{\w x(t)\}_{t> 0}$ only; the calculation of the past path $\{\w x(t)\}_{t\le 0}$ will not be required and will be excluded.

\vspace{0.1cm}
\par
Let $A:\ell_2^+\to \ell_2^+$ be an operator defined as
\baaa
Ay=\Ind_{\ZZ^+}(h\circ y). \eaaa

Consider a mapping $\nu:\ell_2(-\infty,0)\to \ell_2$ such that
 $\nu(x)(t)=x(t)$ for $t\le 0$ and   $\nu(x)(t)=0$ for $t> 0$.

 Let  a mapping $a:\ell_2(-\infty,0)\to \ell_2^+$  be defined   as
\baaa
a(x)=\Ind_{\ZZ^+}\left(h\circ(\nu(x))\right). \eaaa
\par
  Since $h(t)=\O\,\sinc(\O t)/\pi$, the operator  $A$ can be represented  as a matrix with the components \baaa A_{t,m}
=\Ind_{\{t>0,m>0\}}\frac{\O}{\pi}\sinc[\O (t-m)], \quad t,m\in\ZZ,
\label{A}\eaaa
and a process $a(x)=\{a(x,t)\}_{t\in\ZZ}$ can be represented  as a vector  \baaa a(x,t)=\Ind_{\{t>0\}}\frac{\O}{\pi}\sum_{m\le 0} x_m \sinc[\O(t-m)],\quad t\in\ZZ.\label{a}\eaaa
\begin{theorem}
\label{ThM} For any $x\in\ell_2(-\infty,0)$, the equation \baa y=Ay+a(x)
\label{yAa}
\eaa has a unique solution  $\w y(t)=\Ind_{\{t>0\}} \w x(t)\in \ell_2^+$. In addition,  $y=\w x|_{t>0}$, where
$\w x\in \ell_2^\BL$ is defined in Lemma \ref{lemma1}. In other words, $\w y$ is the sought extension  on $\ZZ^+$ of  the
optimal band-limited approximation of the observed sequence $\{x(t)\}_{t\le 0}$.
\end{theorem}

\subsection{Regularized setting}
Let us consider a modification of the original problem (\ref{min})
\baa &&\hbox{Minimize}\quad  \sum_{t\le 0}|
x_\BL(t)-x(t)|^2 +\rho\| x_\BL\|_{\ell_2}^2 \quad\breakk\hbox{over}\quad  x_\BL\in \XN. \label{minR} \eaa
Here  $\rho\ge 0$ is a parameter.

The setting  with $\rho>0$ helps to avoid selection of $\w x$ with an excessive norm.
It can be noted that it is common to
  put restrictions on the norm of the optimal process
  in the data recovery, extrapolation, and interpolation problems in signal processing; see e.g. \citet{Alem,CTao,TH}.

Lemma \ref{lemma1} can be generalized as the following.
\begin{lemma}\label{lemma1R}
For any $\rho\ge 0$ and  $x\in \ell_2(-\infty,0)$, there exists a unique optimal solution  $\w x_\rho$
of the minimization problem (\ref{minR}).
\end{lemma}
\par
In these notations, $\w x_0$ is the optimal process presented in Lemma \ref{lemma1}.\par

\par
Under the assumptions of Lemma \ref{lemma1R}, the
trace on $\ZZ^+$ of the band-limited solution $\w x_\rho$ of problem (\ref{minR})  can
be interpreted  as an optimal forecast of $x|_{\ZZ^-}$ (optimal in the sense of problem (\ref{minR})  given $\O$ and $\rho$).
Let us derive an equation for this solution.
\vspace{0.1cm}
\par Let $I:\ell_2^+\to \ell_2^+$ be the identity operator.
\par It can be noted that Theorem \ref{ThM} does not imply that the operator $(I-A):\ell_2^+\to \ell_2^+$
is invertible, since $a(\cdot):\ell_2(-\infty,0)\to \ell_2^+$ is not a continuous bijection.

 Let
$A_\rho=(1+\rho)^{-1}A$ and $a_\rho(x)=(1+\rho)^{-1}a(x)$, where  $A$  and $a(x)$ are such as  defined  above.

The following lemma shows  that the mapping $A$ is not a contraction but it is close to a contraction, and
$A_\rho$ is a contraction for $\rho>0$.
\begin{lemma}
\label{lemma3}
\begin{enumerate}
\item For any $y\in \ell_2^+$ such that $y\neq 0$,  $\|Ay\|_{\ell_2^+}<\|y\|_{\ell_2^+}$.
\item The operator $A:\ell_2^+\to \ell_2^+$  has the norm $\|A\|= 1$.
\item  For any $\rho\ge 0$, the operator $A_\rho:\ell_2^+\to \ell_2^+$  has the norm $\|A_\rho\|= 1/(1+\rho)<1$.
\item For any $\rho > 0$, the operator $\left(I-A_\rho\right)^{-1}: \ell_2^+\to \ell_2^+$ is continuous and
 $\left\|(I-A_\rho)^{-1}\right\|  \le 1+\rho^{-1}$ for the corresponding norm.
 \end{enumerate}
\end{lemma}
\par
   In addition, by the properties of the projections presented in the definition  for $a(x)$, we have that $\|a_\rho(x)\|_{\ell_2^+}\le \|x\|_{\ell_2(-\infty,0)}$.
\par
Theorem \ref{ThM} stipulates  that equation (\ref{yAa}) has a unique solution.  However, this theorem does not establish
the continuity of the dependence of $\w y$ on the input $x|_{t\le 0}$.
The following theorem shows that additional regularization can be obtained for solution of problem  (\ref{minR}) with $\rho>0$.
 \begin{theorem}
\label{ThMR}  For any $\rho\ge 0$ and $x\in\ell_2(-\infty,0)$, the  equation \baa
(1+\rho)y=Ay+a(x)
\label{yAaR}
\eaa  has a unique solution   $y_\rho=\Ind_{\ZZ^+}\w x_\rho\,=(I-A_\rho)^{-1}a_\rho(x)$ in $\ell_2^+$. Furthermore,
for any $\rho>0$,
\baaa
\|y_\rho\|_{\ell^+_2}\le  \nnn \|x\|_{\ell_2(-\infty,0)}
\label{eest}
\eaaa
for any $x\in\ell_2(-\infty,0)$.
In addition,  $y_\rho=\w x_\rho|_{t>0}$, where
$\w x_\rho\in \ell_2^\BL$ is defined in Lemma \ref{lemma1R}. In other words, $y_\rho$ is the sought extension  on $\ZZ^+$ of  the
optimal band-limited approximation of the observed sequence $\{x(t)\}_{t\le 0}$ (optimal in the sense of problem (\ref{minR}) given $\O$ and $\rho$).
\end{theorem}

Replacement of the original problem by problem (\ref{minR}) with $\rho\to 0$ can be regarded as a Tikhonov regularization of the original problem. By Theorem \ref{ThMR}, it
 leads to solution  featuring continuous dependence on $x|_{t\le 0}$ in the corresponding  $\ell_2$-norm.

\begin{remark}\label{remIT} Since the operator $A_\rho$ is a contraction, the solution of (\ref{yAaR}) can be approximated by  partial sums  $\sum_{k=0}^dA_\rho^ka_\rho(x)$.
\end{remark}

\section{Numerical stability and robustness}
 Let us consider a situation where an input process $x\in\ell_2(0,+\infty)$ is observed with an error.
 In other words, assume that we observe a process $x_\eta=x+\eta$, where $\eta  \in\ell_2(0,+\infty)$ is a noise.
 Let $y_\eta$ be the corresponding solution of equation (\ref{yAaR})  with $x_\eta$ as an input, and let $y$
 be the corresponding solution of equation (\ref{yAaR})  with $x$ as an input. By Theorem \ref{ThMR}, it follows immediately that, for all $\rho>0$ and $\eta\in \ell_2(-\infty,0)$,
  \baaa
  \|y-y_\eta\|_{\ell_2^+}\le \nnn \| \|\eta\|_{\ell_2(-\infty,0)}.
  \eaaa
  This demonstrates some robustness of the method  with respect to  the noise in the observations.

In particular, this ensures robustness with respect to truncation of the input processes,
such that semi-infinite sequences $x\in \ell_2(-\infty,0)$ are replaced by truncated sequences $x_\eta(t)=\Ind_{\{t>q\}}x(t)$ for $q<0$; in this case
   $\eta(t)=\Ind_{t\le q}x(t)$ is such that  $\|\eta\|_{\ell_2(-\infty,0)}\to 0$ as $q\to -\infty$.
 This  overcomes principal impossibility to access infinite sequences of observations.

Furthermore,   only  finite-dimensional systems of linear equations can be solved numerically.
 This means that equation (\ref{yAaR})  with an infinite matrix $A$ cannot be solved exactly even for truncated inputs, since
  it involves  a sequence $a(x)$ that has an infinite support even for truncated  $x$.
   Therefore, we have to apply the method with $A$ replaced by its truncated version.
   We will consider below the impact of  truncation of matrix $A$.
 \subsection*{Robustness with respect to the data errors and truncation}
   Let us  consider replacement  of the matrix  $A=\{A_{t,m}\}_{k,m\in\ZZ^+}$ in equation (\ref{yAaR}) by truncated matrices
  $A_N=\{A_{N,t,m}\}_{t,m\in\ZZ}=\{\Ind_{|t|\le N,|m|\le N}A_{t,m}\}_{t,m\in\ZZ} $ for integers $N>0$. This addresses the restrictions on the data size for numerical
methods.
   Again, we consider a situation where an input process  is observed with an error.
 In other words, we assume that we observe a process $x_\eta=x+\eta \in\ell_2(-\infty,0)$, where $\eta \in\ell_2(-\infty,0)$ is a noise.
 As was mentioned above, this allows to take into account truncation of the inputs as well.

Let us show that the method is robust with  respect to these variations.

Let  $A_{\rho,N}=(1+\rho)^{-1}A_N$.
\begin{lemma}
\label{lemma4} For any $N>0$ , the following holds.
\begin{enumerate}
\item If  $y\in \ell_2^+$ and $\min_{t=1,...,N}|y(t)|>0$, then  $\|A_Ny\|_{\ell_2^+}<\|y\|_{\ell_2^+}$.
\item If  $y\in \ell_2^+$, then  $\|A_Ny\|_{\ell_2^+}\le \|y\|_{\ell_2^+}$.
\item The operator $\left(I-A_N\right)^{-1}: \ell_2^+\to \ell_2^+$ is continuous
 and
 \baaa
 \left\|(I-A_N)^{-1}\right\|<+\infty,
  \eaaa
  for the corresponding norm.
\item For any $\rho>0$, the operator $\left(I-A_{\rho,N}\right)^{-1}: \ell_2^+\to \ell_2^+$ is continuous
 and
 \baaa
 \left\|\left(I-A_{\rho,N}\right)^{-1}\right\|\le 1+\rho^{-1}
  \eaaa
  for the corresponding norm.
   \item For any $\rho\ge 0$ and any $x\in\ell_2(-\infty,0)$, the equation \baa (1+\rho)y=A_Ny+a(x)
\label{eqN}
\eaa has a unique solution  $\w y\in \ell_2^+$.
\end{enumerate}
\end{lemma}
  \begin{theorem}\label{Th3}
 For any $\rho>0$,
    \baaa
    &&\|y_{\rho,\eta,N}-y_\rho\|_{\ell_2^+}\breakk
    \le (1+\rho^{-1})\Bigl(\|(A_N-A)y_\rho \|_{\ell^+_2}+\|\eta \|_{\ell_2(-\infty,0)}\Bigr).
    \eaaa
 Here  $y_\rho$ denote  the solution  in $\ell_2^+$ of equation (\ref{yAaR}), and  $y_{\rho,\eta,N}$ denote the solution in $\ell_2^+$ of equation (\ref{eqN})  with $x$ replaced by $x_\eta=x+\eta$,
 where  $x\in\ell_2(-\infty,0)$ and $\eta\in\ell_2(-\infty,0)$.
 \end{theorem}
 Theorem \ref{Th3} implies robustness with respect to truncation of $(A,x)$ and with respect to the presence of the noise in the input, as the following corollary shows.
  \begin{corollary}\label{corrT}
 For $\rho>0$, solution of equation (\ref{yAaR}) is robust with respect to data errors and truncation, in the sense that
    \baaa
    \|y_{\rho,\eta,N}-y_\rho\|_{\ell_2^+}\to 0\quad \hbox{as}\quad N\to +\infty,\quad\|\eta\|_{\ell_2(-\infty,0)}\to 0 .
    \eaaa
  \end{corollary}
 This justifies acceptance of a result for $(A_N,x_\eta)$ as an approximation of the sought  result for $(A,x)$.
\section{Proofs}
{\em Proof of Proposition \ref{propU}}.
 It suffices to
prove that if $x(\cdot)\in\XN $ is such that $x(t)=0$ for
$t\le 0$, then $x(t)=0$ for $t>0$.
Let $D\defi\{z\in\C: |z|< 1\}$. Let  $H^2(D)$ be the Hardy space of functions that are holomorphic on
$D$ with finite norm
$\|h\|_{\H^2(D)}=\sup_{\rho<1}\|h(\rho e^{i\o})\|_{L_2(-\pi,\pi)}$; see e.g. \cite{Rudin}, Chapter 17.
 It suffices to
prove that if $x(\cdot)\in\XN $ is such that $x(t)=0$ for $t\le 0$,  then
 $x(t)=0$ for $t>0$. Let $X=\Z x$. Since $x\in\XN$, it follows that  $X\in\BNN$.
We have that $X|_D=(\Z x)|_D\in H^2(D)$. Hence, by the property of the Hardy space,
  $X\equiv 0$; see e.g. Theorem 17.18 from \cite{Rudin}.
This
completes the proof of Lemma \ref{propU}. $\Box$
\par
It can be noted that the statement of
Proposition  \ref{propU} can be also derived from predictability of band-limited processes established in  \citet{D12a} or \citet{D12b}.
\par
 {\em
Proof  of Lemma \ref{lemma1}.}  It
suffices to  prove that $\XNL $ is a closed linear subspace of
$\ell_2(-\infty,0)$.  In this case,  there exists a unique projection $\w
x$ of $x|_{\TT}$ on $\XNL$, and the theorem will be
proven.
\par
Consider the mapping $\zeta:\BN \to \XNL$ such that
$x(t)=(\zeta (X))(t)=(\Z^{-1} X)(t)$ for $t\in\TT$. It is a linear
continuous operator. By Proposition \ref{propU}, it is a bijection.
\par
 Since  the mapping $\zeta:\BN \to \XNL$ is continuous, it follows that
the inverse mapping $\zeta^{-1}: \XNL\to\BN$ is also
continuous; see e.g. Corollary in Ch.II.5 \citet{Yosida}, p. 77. Since the
set $\BN$ is a closed linear subspace of $L_2(-\pi,\pi)$, it
follows that $\XNL$ is a closed linear subspace of $\ell_2(-\infty,0)$.
 This completes the proof of  Lemma \ref{lemma1}.  $\Box$
\par
{\em Proof of Theorem \ref{ThM}.}  Let $\w x\in\ell_2^\BL$ be the optimal solution described in Lemma \ref{lemma1}. Let
$\X=\{x\in\ell_2:\  x|_{t>0}=\w x|_{t>0}\}$.
For any $x\in\X$ and $\ww x_\BL\in \ell_2^\BL$, we have that
 \baaa
&& \|\w x-x\|_{\ell_2}^2  =\|\w x-x\|_{\ell_2(-\infty,0)}^2+\|\w x-x\|_{\ell_2(1,+\infty)}^2\\&&=\|\w x- x\|^2_{\ell_2(-\infty,0)}\le \|\ww x_\BL-x\|^2_{\ell_2(-\infty,0)}.
\eaaa
The last inequality here holds because $\w x|_{t\le 0}$  is optimal for problem (\ref{min}).
This implies that, for any $x\in \X$, the sequence   $\w x$
is optimal for  the minimization problem \baaa &&\hbox{Minimize}\quad  \|
x_\BL-x\|_{\ell_2}\quad\hbox{over}\quad x_\BL\in \XN.\quad\label{minP} \eaaa
\par
By the property of the low-pass filters, $\w x= h\circ x$.  Hence the optimal process $\w x\in \XN$ from Lemma \ref{lemma1} is such that
\baaa \w x=h\circ\left(\nu(x)+\Ind_{\ZZ^+}\w x \right). \eaaa
For $\w y=\Ind_{\ZZ^+}\w x$, we have that
\baaa &&\w y=\Ind_{\ZZ^+}\left( h\circ\left(\nu(x)+\Ind_{\ZZ^+}\w x\right)\right)\breakk =
\Ind_{\ZZ^+} (h\circ\nu(x)) +\Ind_{\ZZ^+}(h \circ (\Ind_{\ZZ^+}\w x))\breakk=a(x) +A\w y. \label{xy}\eaaa
This completes the proof  of Theorem \ref{ThM}. $\Box$
\par
{\em Proof  of Lemma \ref{lemma1R}.} As was shown in the proof of Lemma \ref{lemma1},  $\XNL$ is a closed linear subspace
of $\ell_2(-\infty,0)$. The quadratic form in (\ref{minR})  is positive-definite. Then the existence and the uniqueness
of the optimal solution  follows. $\Box$

\par
{\em Proof of Lemma \ref{lemma3}.} Let us prove statement (i). Let $y\in\ell_2^+$. In this case,
$y\notin \XN$;  it follows, for instance, from Proposition \ref{propU}.  Let $Y=\Z y$. We have that $\Z(h\circ y)=H\ew Y\ew$.
Hence   $\|H\ew Y\ew\|_{L_2(-\pi,\pi)}< \|Y\ew\|_{L_2(-\pi,\pi)}$. This implies that  $\|h\circ y\|_{\ell_2}< \|y\|_{\ell_2}$
and that
\baaa
\|A y\|_{\ell_2^+}=\|\Ind_{\ZZ^+}(h\circ y)\|_{\ell_2}\le \|h\circ y\|_{\ell_2}< \|y\|_{\ell_2}=\|y\|_{\ell_2^+}.
\eaaa
This completes the proof of statement (i) of  Lemma \ref{lemma3}.

Let us prove statement (ii). It follows from statement (i)   that  $\|A\|\le 1$. Hence
it  suffices to construct a sequence   $\{y_k\}_{k=1}^{+\infty}\subset\ell_2^+$ such that
\baa
\|A y_k\|_{\ell_2^+}-\|y_k\|_{\ell_2^+}\to 0 \quad \hbox{as}\quad k\to +\infty.
\label{Alim}
\eaa
Let $x\in \XN$ be selected such that $\|x\|_{\ell_2}>0$. Then $h\circ x=x$. Let $x_k$ be defined as $x_k(t)=x(t-k)$, $k\in\ZZ^+$, $k\to +\infty$.
Then  $x_k\in \XN$  and hence   $h\circ x_k=x_k$.
 Let $y_k=\Ind_{\ZZ^+}\, x_k$. By the definitions,
 \baaa
 Ay_k=\Ind_{\ZZ^+}(h\circ (\Ind_{\ZZ^+}\, x_k))= \xi_k+\zeta_k,
 \eaaa
 where
 \baaa
 \xi_k=\Ind_{\ZZ^+}(h\circ x_k), \qquad  \zeta_k=\Ind_{\ZZ^+} (h\circ (\Ind_{\ZZ^+}\, x_k- x_k)).
 \eaaa
Since $h\circ x_k=x_k$, we have that $\xi_k=\Ind_{\ZZ^+}\, x_k=y_k$, i.e. $Ay_k=y_k+\zeta_k$. Further, we have that $
 \zeta_k=-\Ind_{\ZZ^+} (h\circ (\Ind_{\ZZ^-}\, x_k))$.
 Hence
 \baaa
&& \|\zeta_k\|_{\ell_2^+}^2\le  \|h\circ (\Ind_{\ZZ^-}\,x_k)\|_{\ell_2}^2\le \|\Ind_{\ZZ^-}\, x_k\|_{\ell_2}^2\breakk=\sum_{t\le 0}|x_k(t)|^2
 =\sum_{t\le -k}|x(t)|^2\to 0
\eaaa
as $k\to +\infty$. Hence (\ref{Alim}) holds.
This completes the proof of statement (ii) and  Lemma \ref{lemma3}.
\par
Statement (iii)  follows immediately from statement (ii). Statement (iv) follows from the estimates
 \baa
 \left\|(I-A_\rho)^{-1}\right\| \le \sum_{k=0}^\infty \|A_\rho\|^k=\frac{1}{1-[\|A\|/(1+\rho)]^k}
 \brea=1/(1-1/(1+\rho))=1+\rho^{-1}.\quad\quad\qquad
  \label{1+rho}\eaa
This completes the proof of statement  Lemma \ref{lemma3}.
$\Box$

\par
{\em Proof of Theorem \ref{ThMR}.}   This proof represents a generalization of the proof
of Theorem \ref{ThM} which covers a special case where $\rho=0$.

Let $\w x_\rho\in\ell_2^\BL$ be the optimal solution described in Lemma \ref{lemma1R}. Let
$\X_\rho=\{x\in\ell_2:\  x|_{t>0}=\w x_\rho|_{t>0}\}$.
For any $x\in\X_\rho$ and $\ww x_\BL\in \ell_2^\BL$, we have that
 \baaa
&& \|\w x_\rho-x\|_{\ell_2}^2+ \rho\|\w x_\rho\|_{\ell_2}^2\breakk  =\|\w x_\rho- x\|_{\ell_2(-\infty,0)}^2+\|\w x_\rho- x\|_{\ell_2(1,+\infty)}^2+ \rho\|\w x_\rho\|_{\ell_2}^2
\\&&=\|\w x_\rho- x\|^2_{\ell_2(-\infty,0)}+ \rho\|\w x_\rho\|_{\ell_2}^2\breakk\le \|\ww x_\BL- x\|^2_{\ell_2(-\infty,0)}+ \rho\|\ww x_\BL\|_{\ell_2}^2.
\eaaa
The last inequality here holds because the path $\w x_\rho|_{t\le 0}$ is optimal for problem (\ref{minR}).
This implies that, for any $x\in \X_\rho$, the sequence   $\w x_\rho$
is optimal for  the minimization problem \baaa &&\hbox{Minimize}\quad  \|
x_\BL-x\|_{\ell_2}^2+\rho\|x_\BL\|_{\ell_2}^2\quad\breakk\hbox{over}\quad x_\BL\in \XN.\quad\label{minPP} \eaaa

Let us show  that
\baa \w x_\rho=\frac{1}{1+\rho}\, h\circ\left(\nu(x)+\Ind_{\ZZ^+}\w x_\rho\right). \label{R}\eaa
\par
Let $x\in\ell_2$ and $x_\rho'=\Ind_{\ZZ^-}x+\Ind_{\ZZ^+}\w x_\rho$.  Since $x_\rho'\in\X_\rho$, it follows  that $\w x_\rho$
is an unique solution of  the minimization problem \baaa &&\hbox{Minimize}\quad  \|
x_\BL-x_\rho'\|_{\ell_2}^2+\rho\|x_\BL\|_{\ell_2}^2 \quad\breakk\hbox{over}\quad  x_\BL\in \XN.\label{minRPP} \eaaa

Further, the quadratic form here can be represented
as
\baaa    &&\|x_\BL-x_\rho'\|_{\ell_2}^2+\rho\|x_\BL\|_{\ell_2}^2\breakk=   (1+\rho)(x_\BL,x_\BL)_{\ell_2}-2(x_\BL,x_\rho')_{\ell_2}+
(x_\rho',x_\rho')_{\ell_2}\\&&= (1+\rho)\Bigl[(x_\BL,x_\BL)_{\ell_2} -2\Bigl(x_\BL, \frac{1}{1+\rho} x_\rho'\Bigr)_{\ell_2} \breakk +
\frac{1}{1+\rho}(x_\rho',x_\rho')_{\ell_2}\Bigr]\\&&
= (1+\rho)\Bigl[\Bigl\|x_\BL-\frac{1}{1+\rho} x_\rho'\Bigr\|_{\ell_2}^2-\frac{1}{(1+\rho)^2}(x_\rho',x_\rho')_{\ell_2} \breakk
+\frac{1}{1+\rho}(x_\rho',x_\rho')_{\ell_2}\Bigr].  \eaaa
It follows that   $\w x_\rho=(1+\rho)^{-1}\w x_\rho'$, where   $\w x_\rho'$
is an unique solution of  the minimization problem \baaa &&\hbox{Minimize}\quad  \|
x_\BL-x'_\rho\|_{\ell_2}^2 \quad\hbox{over}\quad x_\BL\in \XN.\label{minR1} \eaaa
By the property of the low-pass filters, $\w x_\rho'=h\circ x_\rho'$.
It follows from the definitions that
\baaa (1+\rho)\w x_\rho= \w x_\rho'=h\circ\left(\nu(x)+\Ind_{\ZZ^+}x'_\rho\right)\brea=h\circ\left(\nu(x)+\Ind_{\ZZ^+}\w x_\rho\right). \eaaa
This proves  (\ref{R}).
\par
Further, equation (\ref{R}) is  equivalent to equation (\ref{yAaR}) which, on its turn, is  equivalent to the equation  \baaa y=A_\rho y+a_\rho (x).
\label{yR2}
\eaaa
Since the operator $(I-A_\rho)^{-1}: \ell_2^+\to \ell_2^+$ is continuous, this equation  has an unique solution
$y_\rho=\Ind_{\ZZ^+}\w x_\rho=(I-A_\rho)^{-1}a_\rho(x)$ in $\ell_2^+$, and the required estimate for $\|y_\rho\|_{\ell_2^+}$ holds.
 This completes the proof of Theorem \ref{ThMR}. $\Box$
\par
{\em Proof of Lemma \ref{lemma4}.} Let us prove statement (i). The proof follows the approach of the proof of Lemma \ref{lemma3}(i).
Let $D_N=\{1,2,...,N\}$, and let $z=\Ind_{D_N}y\in\ell_2^+$.
Under the assumptions on $y$, we have that $z\neq 0$. In this case,
$z\notin \XN$;  it follows, for instance, from Proposition \ref{propU}.  Let $Z=\Z z$. We have that $\Z(h\circ z)=H\ew Z\ew$.
Hence   $\|H\ew Z\ew\|_{L_2(-\pi,\pi)}< \|Z\ew\|_{L_2(-\pi,\pi)}$. This implies that  $\|h\circ z\|_{\ell_2}< \|z\|_{\ell_2}$.
Hence
\baaa
\|A_N y\|_{\ell_2^+}=\|\Ind_{D_N}(h\circ z)\|_{\ell_2}\le \|h\circ z\|_{\ell_2}< \|z\|_{\ell_2}\le \|y\|_{\ell_2^+}.
\eaaa
This completes the proof of statement (i). The proof of (ii) is similar; in this case, the case where $z=0$ is not excluded.

 Let us prove statements (iii). Consider a matrix $\oo A_N=\{A_{t,m}\}_{1\le t,m\le N}\in\R^{N\times  N}$. Let $I_N$ be the unit matrix in $\R^{N\times N}$.
 Suppose that the matrix $I_N-\oo A_N$ is degenerate, i.e. that there exists a non-zero $z=\{z(t)\}_{t=1}^N\in\R^N$ such that $\oo A_N z=z$.
Let $y\in\ell_2^+$ be such that $y(t)=\Ind_{1\le t\le N}z(t)$.
In this case, $A_N y=y$ which  would contradict  the statement (i).
  Therefore, the matrix $I_N-\oo A_N$  is non-degenerate. Hence the operator $(I_N-\oo A_N)^{-1}:\R^N\to\R^N$
  is continuous and
    $\|(I_N-\oo A_N)^{-1}\|<+\infty$ for the corresponding norm.
\par
  The space $\ell_2^+$ is isomorphic to the space  $\Y=\R^N\times \ell_2(N+1,+\infty)$, i.e. $y\in \ell_2^+$ can be represented as
   $(\oo y, \ww y)\in \Y$, where  $\oo y=(y(1),...,y(N))^\top\in \R^N$ and $\ww y=y|_{t>N}\in \ell_2(N+1,+\infty)$. Respectively,
   the sequence $A_N y\in\ell_2^+$ can be represented as $(\oo A_N \oo y, 0_{\ell_2(N+1,+\infty)} )\in\Y$, and the sequence $y-A_Ny\in\ell_2^+$ can be represented as $(\oo y-\oo A_N \oo y, \ww y)\in\Y$.
   Hence the sequence $(I_N-A_N)^{-1}y\in\ell_2^+$ can be represented as $((I_N-\oo A_N)^{-1}\oo y, \ww y|_{t>N})\in\Y$. Clearly,
   \baaa
  \|(I-A_N)^{-1}y\|_{\ell_2^+}^2\le |(I_N-\oo A_N)^{-1}\oo y|^2+\|\ww y\|_{\ell_2(N+1,+\infty)}^2
   \brea \le \|(I_N-\oo A_N)^{-1}\|^2 |\oo y|^2+\|\ww y\|_{\ell_2(N+1,+\infty)}^2.
\eaaa
This proves statement (iii).
\par
The proof of statement (iv) repeats estimates (\ref{1+rho}) if we take into account  that $\|A_N\|\le \|A\|= 1$.
\par
  To complete the proof of Lemma \ref{lemma4}, it suffices to observe that
statement  (v)  for $\rho=0$ follows from statement
(iii), and  statement  (v) for $\rho>0$  follows from statement
(iv).  $\Box$
\par
 {\em Proof of Theorem \ref{Th3}}.
  Let $e_N=y_{\rho,\eta,N}-y_\rho$.  We have that
     \baaa (1+\rho)e_N=A_Ne_N+ (A_N-A)y_\rho+a(x_\eta)-a(x).
   \label{AAN}  \eaaa
   By the properties of the sinc functions presented in
     (\ref{a}), it follows that \baaa
     \|a(x)-a(x_\eta)\|_{\ell_2^+}\le \|\eta\|_{\ell_2(-\infty,0)}.
     \eaaa
    Hence  \baaa
 && \|e_{\rho,\eta,N}\|_{\ell^+_2}\breakk \le \|(I-A_{\rho,N})^{-1}\|\Bigl[\|(A_N-A)y_\rho +a_\rho(x)-a_\rho(x_\eta)\|_{\ell^+_2}\Bigr]
   \\ &&\le \|(I-A_{\rho,N})^{-1}\|\Bigl[\|(A_N-A)y_\rho \|_{\ell^+_2}+\|\eta \|_{\ell_2(-\infty,0)}\Bigr].\eaaa
This completes the proof of Theorem \ref{Th3}. $\Box$

    {\em Proof of  Corollary \ref{corrT}}.
  We have that $A_N y=\Ind_{D_N}(h \circ (\Ind_{D_N}y))$, where $D_N=\{1,2,...,N\}$.
    Hence
   \baaa
   (A_N-A)y_\rho= \Ind_{D_N}(h \circ (\Ind_{D_N}y_\rho ))-\Ind_{\ZZ^+}(h \circ y_\rho ))\brea=\w\zeta_{N,\rho}+\ww \zeta_{N,\rho},
   \eaaa
   where
    \baaa
   \w\zeta_{N,\rho}= \Ind_{D_N}[h \circ (\Ind_{D_N}y_\rho )-h \circ y_\rho ]=\Ind_{D_N}[h \circ (\Ind_{D_N}y_\rho - y_\rho) ]\\=
   \Ind_{D_N}[h \circ (\Ind_{D_N}y_\rho - y_\rho) ]=-\Ind_{D_N}[h \circ (\Ind_{\{t:\ t>N\}}y_\rho)\eaaa
   and \baaa
   \ww\zeta_{N,\rho}= [\Ind_{D_N}-\Ind_{\ZZ^+}](h \circ y_\rho )= -\Ind_{\{t:\ t> N\}}(h \circ y_\rho ).
   \eaaa
   Clearly, $\|\w\zeta_{N,\rho}\|_{\ell_2^+}\to 0$ and  $\|\ww\zeta_{N,\rho}\|_{\ell_2^+}\to 0$ as $N\to +\infty$. Hence $\|(A_N-A)y_\rho\|_{\ell_2^+}\to 0$ as $N\to +\infty$.
 This completes the proof of  Corollary \ref{corrT}. $\Box$

\section{Some numerical experiments}\label{secN}
We did some  numerical  experiments to compare statistically the performance of our band-limited  extrapolations with
extrapolations based on splines applied to causally smoothed processes. In addition, we
did some  numerical  experiments to estimate  statistically  the impact of data truncation.
\subsection{Simulation of the input processes}
       The setting of Theorems \ref{ThM}-\ref{ThMR}  does not involve stochastic processes and probability measure; it is oriented on extrapolation of sequences
       in the pathwise deterministic setting.
      However, to  provide  sufficiently large  sets of input sequences for statistical estimation,  we used processes $x$ generated via Monte-Carlo simulation
      as a stochastic  process evolving as
      \baa
       &&z(t)={\rm A}(t) z(t-1)+\eta(t), \quad  t\in\ZZ,\qquad \breakk  x(t)=c^\top z(t).
       \label{AR2}\eaa
       Here $z(t)$ is a process with the values in $\R^\nu$, where $\nu\ge 1$ is an integer, $c\in \R^\nu$.
       The process  $\eta$ represents a noise  with values in $\R^\nu$,
          ${\rm A}(t)$ is a matrix with the values in
        $\R^{\nu\times \nu}$ with the spectrum inside $\T$.
        The matrices ${\rm A}(t)$ are switching  values randomly at random times; this replicates a  situation where the parameters of a system
        cannot be recovered from the observations such as described in the review \cite{SFS}.

  Since it is impossible to implement Theorem \ref{ThMR} with infinite input sequences,
one has to use truncated  inputs for  calculations.
In the experiments described below, we replaced  $A$ and $x|_{t\le 0}$ by their
truncated analogs
\baaa
A_N=\{\Ind_{\{|k|\le N, |m|\le N\}}A_{k,m}\},\quad x_N=\Ind_{\{t\ge -N\}}x(t),\quad
\label{N}
\eaaa where  $N>0$ is the truncation horizon.

In each simulation, we selected random and mutually independent $z(-N)$, ${\rm A}(\cdot)$, and $\eta$, as vectors and
 matrices with  mutually independent components.
   The process $\eta$ was selected as a stochastic discrete time Gaussian white noise with the values in $\R^\nu$
such that $\E \eta(t)=0$ and $\E |\eta(t)|^2=1$. The initial vector  $z(-N)$ was selected randomly  with the components
     from the uniform distribution on $(0,1)$.
   The components of the matrix ${\rm A}(-N)$ was selected from the uniform distribution on $(0,1/\nu)$. Further, to simulate randomly changing ${\rm A}(t)$,
     a random variable $\xi$ distributed
     uniformly on $(0,1)$ and  independent on $({\rm A}(s)|_{s<t},\eta,z(-N))$ was simulated for each time $t>-N$.
         In the case where $\xi<0.5$,  we selected
     ${\rm A}(t)={\rm A}(t-1)$. In the case where $\xi\ge 0.5$,
     ${\rm A}(t)$ was simulated randomly  from the same distribution as ${\rm A}(-N)$, independently on $({\rm A}(s)|_{s<t},\eta,z(-N))$.
This setting  with randomly changing  ${\rm A}(t)$ makes impossible to identify the parameters of equation (\ref{AR2})  from the current observations.

In our experiments, we calculated
the solution $\w x_\rho|_{t>0}$ of linear system (\ref{yAaR}) for a given $x$ directly using a built-in MATLAB operation for solution of
linear algebraic
systems.
\subsection{Comparison with spline extrapolations}
We compared the accuracy of the band-limited extrapolations introduced in Theorem \ref{ThMR} with the accuracy of
three standard extrapolations  built in MATLAB:  piecewise cubic spline extrapolation,
shape-preserving piecewise cubic extrapolation, and linear extrapolation.

We denote by $\EE$ the sample mean across the Monte Carlo trials.

We estimate the values
    \baaa
      e_\BL=\EE\sqrt{\sum_{t=1}^L|x(t)-\w x_\BL(t)|^2},\label{err}\eaaa
where  $\w x_\BL$ is an extrapolation calculated as suggested in Theorem \ref{ThMR} with some $\rho>0$, i.e.
$\w x_\BL |_{t>0}=y_\rho=\w x_\rho|_{t>0}$, in the terms of this theorem,
 for some integers $L>0$. The choice of $L$ defines the extrapolation horizon; in particular, it defines prediction horizon if extrapolation is used for forecasting.

 We compare these values with similar values obtained for some standard spline extrapolations
 of the causal $h$-step moving average process for $x$. More precisely, to
 take into the account truncation, we used a modification of the causal moving average \baaa
 \oo x(t)=\frac{1}{\min(h,t+N+1)}\sum_{k=\max(t-h,-N)}^t x(k),\quad t\ge -N.
 \eaaa
 \par
 For three selected  standard spline extrapolations, we calculated
  \baaa
      e_d=\EE\sqrt{\sum_{t=1}^L|x(t)-\ww x_d(t)|^2},\quad d=1,2,3, \label{err2}\eaaa
 where $\ww x_1$ is
 the  piecewise cubic extrapolation  of the moving average $\oo x|_{t\le 0}$,
 $\ww x_2$  is
 the shape-preserving piecewise cubic extrapolation of $\oo x|_{t\le 0}$,  $\ww x_3$ is
 the linear  extrapolation of $\oo x|_{t\le 0}$.

 We used these extrapolation applied to the moving average since
 applications directly to the process $x(t)$
produce  quite unsustainable extrapolation with  large values $e_d$.

\par

We calculated and compared $e_\BS$ and $ e_d$, $d=1,2,3$.
Table \ref{simulation} shows the ratios  $e_{\BL}/e_d$ for some combinations of parameters. For these calculations, we used  $c=(1/\nu,1/\nu,...,1/\nu)^\top$, $h=10$, and $\rho=0.4$.
\begin{table}[h]
\caption{Comparison of performance of band-limited extrapolation and standard extrapolations. }
  \centering
\begin{tabular}{lcccc}
  \hline
& &$e_{\BL}/ e_1$
&$e_{\BL}/ e_2$
&$e_{\BL}/ e_3$  \\

  \hline \\
&\multicolumn{4}{c}{Panel (a): $\nu=1$, $\O=\pi/2$, $N=50$} \\
$L=1$   && 0.8818 & 0.9312 & 0.9205   \\
$L=3$   && 0.4069 & 0.8407 & 0.9270   \\
$L=6$   && 0.1017 & 0.3095 & 0.8330   \\
$L=12$   && 0.0197 & 0.0489 & 0.6751    \\
\\
&\multicolumn{4}{c}{Panel (b): $\nu=8$, $\O=\pi/5$, $N=100$} \\
$L=1$   && 0.9255 & 0.9801 & 0.9633 \\
$L=3$   && 0.3975 & 0.8369 & 0.9348   \\
$L=6$   && 0.1020 & 0.2947 & 0.8426   \\
$L=12$   && 0.0188 & 0.0451 & 0.6739    \\  \hline
\end{tabular}

\index{\vspace{2mm}
{\sm $\EE e_d$ represent the average distances  (\ref{err2}) from the future process:  $\EE e_1$ for
 the  piecewise cubic extrapolation $\ww x_2$ of the moving average, $\EE e_1$ for the shape-preserving piecewise cubic extrapolation,
 $\EE e_3$ for the linear extrapolation; $L$ is the extrapolation horizon.}}
 \label{simulation}   \end{table}

For each  entry in Table \ref{simulation}, we used  10,000
     Monte-Carlo trials.
     The values $e_d$ were calculated
using Matlab programm {\em interp1}. An experiment with 10,000 Monte-Carlo trials would take
      about one minute of calculation time for a standard personal computer. The experiments demonstrated a good numerical stability of the  method;  the results were quite robust with respect to truncation of the input processes and deviations of parameters.
Increasing the number of Monte-Carlo trials gives very close results.
\par
  In addition, we found that the choice of the dimension $\nu$ does not affect much the result.
For example, we obtained  $e_{\BL}/ e_1=0.4069 $ for $L=3$, $\nu=1$, $\O=\pi/2$, $N=50$. When we repeated this experiment with $\nu=8$,
 we obtained  $ e_{\BL}/ e_1=0.4091$
 which is not much different. When we repeated the same experiment  with $\nu=8$ and with 30,000 trials, we obtained  $ e_{\BL}/ e_1= 0.4055$ which is not much different again.

The ratios   $ e_{\BL}/ e_d$ are decreasing further as the horizon $L$ is increasing, hence we omitted the results for $L>12$.  Nevertheless,  the results for large $L$ are not particularly
meaningful since the noise nullifies for large $L$ the value of information collected from observation of
$x|_{t\le 0}$.  We also omitted results with classical extrapolations applied directly to $x(t)$ instead of the moving average $\oo x(t)$, since
errors $e_\BL$ and $e_d$   are quite large in this case due the presence of the noise.

Table \ref{simulation} shows that the band-limited extrapolation performs better than the spline extrapolations;
some additional experiments with other choices of parameters demonstrated the same trend.
However, experiments did not involve more advanced methods beyond the listed above spline methods.
Nevertheless, regardless of the results of these experiments,
 potential importance of band-limited extrapolation is self-evident  because  its
physical meaning:  a band-limited part can be considered
 as a regular part of a process purified from a noise represented by high-frequency component. This is controlled
 by the choice of the band. On the other hand, the choice of particular splines does not have a physical
 interpretation.   \par
\begin{figure}[ht]
\centerline{\psfig{figure=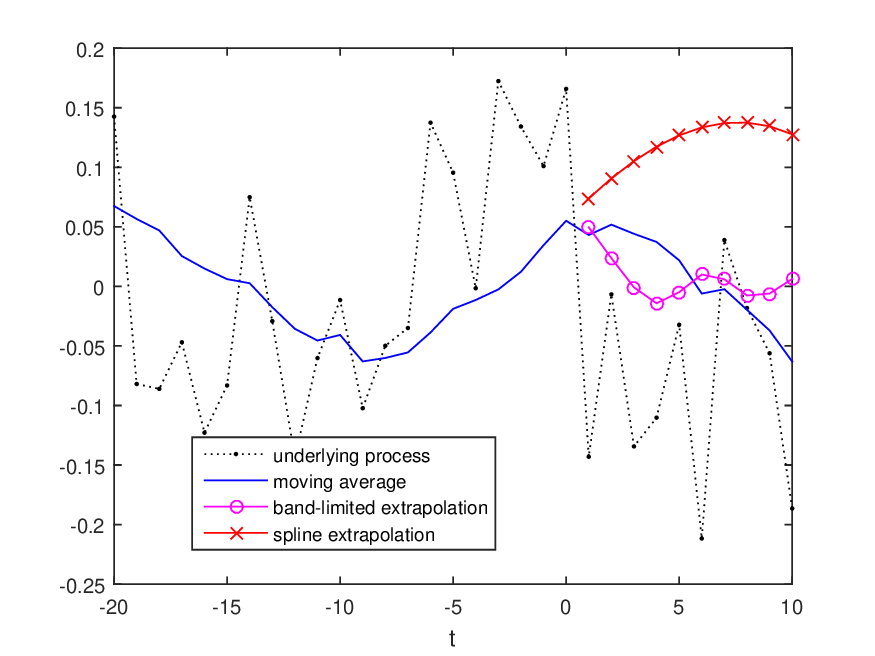,width=9cm,height=7.1cm}}
\caption[]{\sm
Example of a path $x(t)$, its band-limited extrapolation,
its moving  average, and piecewise cubic extrapolation  with
$\O=\pi/2$, $N=50$, $h=10$, $\rho=0.2$. }
\vspace{0cm}\label{fig-1}\end{figure}
\begin{figure}[ht]
\centerline{\psfig{figure=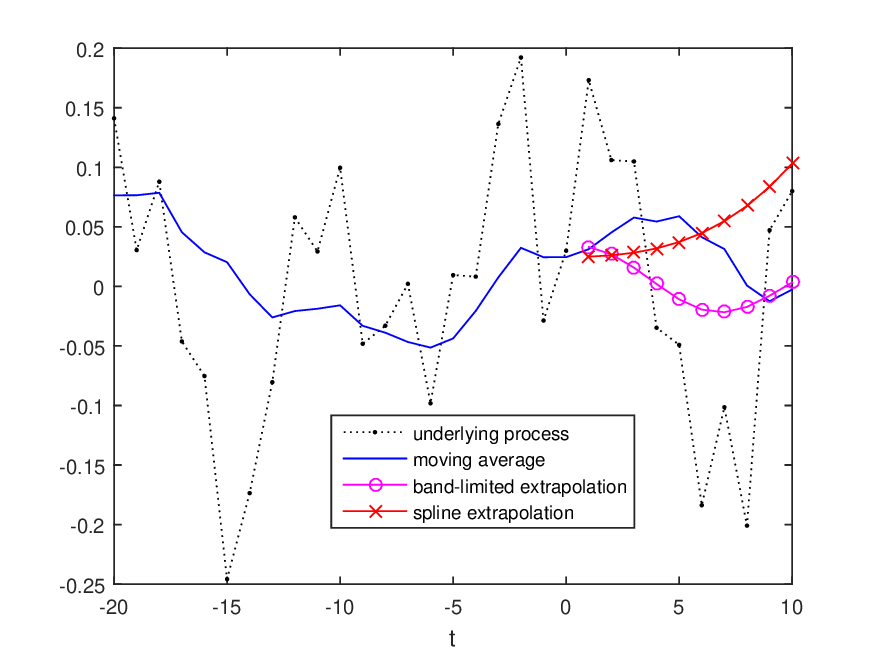,width=9cm,height=7.0cm}}
\caption[]{\sm Example of a path $x(t)$, its band-limited extrapolation,
its moving  average, and shape-preserving piecewise cubic extrapolation  with $\nu=8$,
$\O=\pi/5$, $N=100$, $h=10$, and $\rho=0.4$. } \vspace{0cm}\label{fig-2}\end{figure}

Figures \ref{fig-1} and \ref{fig-2} show examples of paths of processes $x(t)$  plotted against time $t$, their band-limited extrapolations $x_\BL(t)$,
their moving  averages $\oo x(t)$, and their spline extrapolations  $\ww x_k(t)$, $k=1,2$,  with
$\nu=8$,  $h=10$,  $L=10$, and $c=(1,1,..,1)^\top$. Figure \ref{fig-1} shows
piecewise cubic extrapolation  $\ww x_1(t)$, with parameters   $\O=\pi/2$,  $N=50$, $\rho=0.2$.
Figure \ref{fig-2} shows shape-preserving
piecewise cubic extrapolation  $\ww x_2(t)$,  with parameters  $\O=\pi/5$,  $N=100$, $\rho=0.4$.

It can be noted that, since  our method does not require to calculate
 $\w x(t)|_{t\le 0}$, these sequences  were not calculated and are absent on Figures \ref{fig-1}-\ref{fig-2}; the
extension $\w x(t)|_{t>0}$ was derived directly from  $x(t)|_{t\le 0}$.

\subsection{Estimation of the impact of data truncation}
In addition, we   did experiments to estimate the impact of truncation
for the band-limited extrapolations  introduced in Theorem \ref{ThMR}.  We found that
impact of truncation is manageable; it decreases
 if the size of the sample increasing.  In these experiments,  we calculated and compared the values
\baaa
      \!E_{\None,\Ntwo}\!=\!\EE\! \left[\frac{2\sqrt{\sum_{t=1}^L|\w x_{\BL,\None}(t)-\w x_{\BL,\Ntwo}(t)|^2}}{\sqrt{\sum_{t=1}^L\w x_{\BL,\None} (t)^2}+\sqrt{\sum_{t=1}^L\w x_{\BL,\Ntwo} (t)^2}}\right]
\label{errN}\eaaa
      describing the impact of the replacement a truncation horizon  $N=N_1$ by another truncation horizon  $N=N_2$. Here  $\w x_{\BL,\NS}$ is the band-limited extrapolation
calculated with truncated data defined by (\ref{N}) with a truncation horizon $N$; $\EE$ denotes again the average over Monte-Carlo experiments.

We used  $x(t)$ simulated via (\ref{AR2}) with randomly switching ${\rm A}(t)$, the same as in the experiments described  above,
with the following adjustment
for calculation of $E_{\None,\Ntwo}$.
For the case where  $N_2>N_1$, we simulated first  a path  $x|_{t=-N_2,..., 0}$  using equation (\ref{AR2}) with
a randomly selected initial value for $z(-N)$  selected at $N=N_2$ as was described above, and then used the truncated part
$x|_{t=-N_1,..., 0}$
of this path  to calculate $\w x_{\BL,\None}$; respectively, the  path  $x|_{t=-N_2,..., 0}$
was used to calculate $\w x_{\BL,\Ntwo}$.

Table \ref{table2} shows the results of simulations with 10,000 Monte-Carlo trials for each entry  and with
 $\nu=8$, $c=(1,...,1)^\top$, $\O=\pi/2$, $\rho=0.4$, $L=12$.
\index{ \baaa
 E_{50}=  ,\quad   E_{250}=,\quad  E_{500}= 0.
 \eaaa}
\begin{table}[h]
\caption{Impact of the truncation and the choice of the truncation horizon}
  \centering
\begin{tabular}{ccccc}
  \hline $E_{25,50}$
&$E_{50,100}$
&$E_{100,250}$&$E_{250,500}$
&$E_{500,1000}$
\\
  \hline \\
0.0525& 0.0383&  0.0303& 0.0180& 0.0128
\\
\hline
\end{tabular}
\label{table2}
\end{table}
\par
\begin{figure}[ht]
\centerline{\psfig{figure=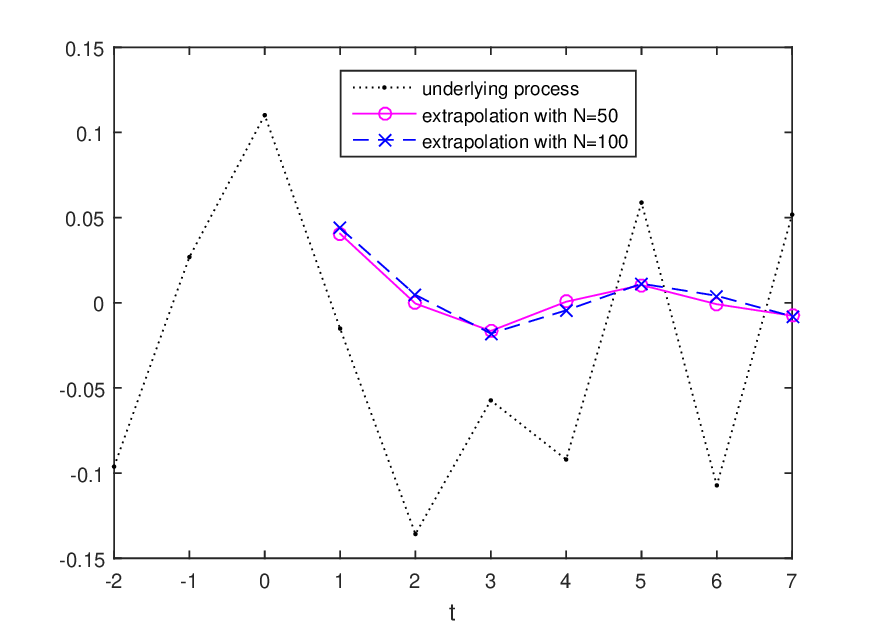,width=9cm,height=7.1cm}}
\caption[]{\sm
Example of a path $x(t)$  and its band-limited extrapolations $\w x_{\BL,N}(t)$
calculated with truncation horizons $N=50$ and $N=100$  in (\ref{N}). }
\vspace{0cm}\label{figT}\end{figure}

Figure \ref{figT} illustrates the results presented in Table \ref{table2} and shows an example of a path $x(t)$  plotted against time $t$ together with the path of its band-limited extrapolations $\w x_{\BL,N}(t)$  obtained
with the same  parameters as for the Table \ref{table2}, with the truncation horizons $N=50$ and $N=100$. The figure shows that the impact of doubling the truncation horizon is quite small, since the paths for extrapolations are quite close.
\section{Discussion and future development}
The  paper suggests a linear equation in the time domain  for calculation of band-limited extensions on the future times of band-limited approximations of one-sided semi-infinite sequences representing past observations
(i.e. discrete time processes in deterministic setting). The method allows to exclude
analysis of processes in the frequency domain and calculation of band-limited approximation of the observed past. This helps to streamline
the calculations.
Some numerical stability and robustness with respect to input errors and
data truncation are established.

It appears that the extrapolation error caused by the truncation is manageable for a short extrapolation horizon  and
can be significant on a long extrapolation horizon, i.e. for large $t>0$. This is because the components $((A_N-A)y_\rho)(t)$ of the input term in (\ref{AAN}) are relatively small for small  $t>0$ and can be large for large $t>0$. In particular, this means that
 long horizon prediction based on this method  will not be particularly efficient.

  There are possible modifications that we leave for the future research.

In particular, the  suggested method  can be extended on the setting where $x(t)$ is
approximated by a "high frequency" band-limited  processes $\w x(t)$ such that the
process $\w X\ew$ is supported on $[-\pi,-\pi+\O]\cup [\pi-\O,\pi]$.
In this case, the solution follows immediately from the solution given above  with
$x(t)$ replaced by $(-1)^t x(t)$. In addition, processes with more general types of the spectrum gaps on $\T$ can be considered, given some  modification of the algorithm.

It could be interesting to see if the estimate in Lemma 4 (iv)
 can be improved; the statement in Lemma 4 (iii) gives a hint that this is estimate is not  sharp  for preselected $N$.

It could be interesting to apply an iteration method similar to the one used in \citet{Zhao2}; see Lemma 1 \citet{Zhao2} and citations therein.
\subsection*{Acknowledgment}
The author would like thank the anonymous reviewers for
the detailed suggestions that improved the manuscript.

\end{document}